\documentclass[a4paper,12pt]{article}
\usepackage{jheppub,esint,psfrag}
\usepackage[utf8]{inputenc}

\usepackage{epsfig,amssymb,amsmath,psfrag,subfigure,rotate,color,wasysym,xcolor}
\usepackage[bbgreekl]{mathbbol}

\allowdisplaybreaks 

\DeclareSymbolFontAlphabet{\mathbbm}{bbold}
\DeclareSymbolFontAlphabet{\mathbb}{AMSb}%

\def\XXint#1#2#3{{\setbox0=\hbox{$#1{#2#3}{\int}$ }
\vcenter{\hbox{$#2#3$ }}\kern-.6\wd0}}

\def \be  {\begin{equation}}
\def \ee  {\end{equation}}
\def \ba  {\begin{eqnarray}}
\def \ea  {\end{eqnarray}}
\def \baa {\begin{eqnarray*}}
\def \eaa {\end{eqnarray*}}

\def \lab #1 {\label{#1}}

\newcommand\re[1]{(\ref{#1})}
\def\d{\hbox{{d}\kern-.20em\hbox{l}}}

\def \matrix #1 {\left(\begin{array}{cc} #1 \end{array}\right)}

\newcommand \vev [1] {\langle{#1}\rangle}

\newcommand \ket [1] {|{#1}\rangle}
\newcommand \bra [1] {\langle {#1}|}

\def\1{\hbox{{1}\kern-.25em\hbox{l}}}

\newcommand{\ft}[2]{{\textstyle\frac{#1}{#2}}}
\makeatletter

\input pdf-trans
\newbox\qbox
\def\usecolor#1{\csname\string\color@#1\endcsname\space}
\newcommand\bordercolor[1]{\colsplit{1}{#1}}
\newcommand\fillcolor[1]{\colsplit{0}{#1}}
\newcommand\outline[1]{\leavevmode%
  \def\maltext{#1}%
  \setbox\qbox=\hbox{\maltext}%
  \boxgs{Q q 2 Tr \thickness\space w \fillcol\space \bordercol\space}{}%
  \copy\qbox%
}
\makeatother
\newcommand\colsplit[2]{\colorlet{tmpcolor}{#2}\edef\tmp{\usecolor{tmpcolor}}%
  \def\tmpB{}\expandafter\colsplithelp\tmp\relax%
  \ifnum0=#1\relax\edef\fillcol{\tmpB}\else\edef\bordercol{\tmpC}\fi}
\def\colsplithelp#1#2 #3\relax{%
  \edef\tmpB{\tmpB#1#2 }%
  \ifnum `#1>`9\relax\def\tmpC{#3}\else\colsplithelp#3\relax\fi
}
\bordercolor{black}
\fillcolor{white}
\def\thickness{.3}

\def\1{\mathbbm{1}}

\title{Collinear bootstrap for $\mathcal{N}=(1,1)$ sYM}

\author[a]{A.V.~Belitsky,}
\author[b]{V.A. Smirnov}
\affiliation[a] {Department of Physics, Arizona State University,  Tempe, AZ 85287-1504, USA}  
\affiliation[b]{Skobeltsyn Institute of Nuclear Physics, Moscow State University 119992 Moscow, Russia}
    
 \abstract
{We study the multi-collinear behavior of tree amplitudes in the six-dimensional $\mathcal{N} = (1,1)$ super Yang-Mills theory (sYM). A 
generalized dimensional reduction of the latter yields the four-dimensional $\mathcal{N} = 4$ sYM on the Coulomb branch, which is 
of interest for considerations of massive or off-shell scattering. To this end, we revisit the calculation of tree scattering in the former 
theory employing the collinear bootstrap and known massless limits. Assuming the universality of the double-collinear asymptotics, 
the result for six-leg superamplitudes differs from the one available in the literature. We further extract the triple-collinear splitting 
superamplitudes from these.}

\begin{document}

\maketitle
\flushbottom
\setcounter{footnote} 0

\section{Introduction}

Scattering amplitudes develop singularities when adjacent external momenta become col\-linear 
\cite{Parke:1986gb,Berends:1987me,Mangano:1990by}. This behavior is universal and depends
solely on the attributes of coalescing particles\footnote{To be precise, this holds in the time-like kinematics only \cite{Catani:2011st}. 
We will assume it throughout our current consideration.}. Its understanding is of paramount importance for phenomenological 
applications, where cancellations between real and virtual divergences ensure finite physical cross sections, on the one hand. 
On the other, known factorization patterns furnish strong constraints on the structure of scattering amplitudes and provide an 
independent check on the correctness of their calculation. 

Presently, we will address this question within the context of the six-dimensional $\mathcal{N} = (1,1)$ super Yang-Mills theory (sYM).
This theory is maximally supersymmetric and serves as a progenitor for $\mathcal{N} = 4$ sYM in four dimensions upon dimensional 
reduction. Depending on whether one keeps out-of-four-dimensional components of particles' six-dimensional null momenta vanishing 
or not, one can probe different branches of the theory. When they are set to zero, one obtains the conventional superconformal model,
while, if they are not, one ends up on its Coulomb branch, where some of the states become massive. Our interest in the past few years 
was in a particular pattern of dimensional reduction that endows only external particles with nontrivial masses while keeping all states 
propagating in quantum loops massless. This can then be regarded as an off-shell generalization of massless on-shell scattering.

The on-shell scattering in the six-dimensional theory enjoys a superspace formulation \cite{Dennen:2009vk}, on the one hand, and 
an unconstrained spinor-helicity formalism \cite{Cheung:2009dc}, on the other.  In addition, amplitudes possess covariant transformation 
properties under the dual conformal symmetry \cite{Bern:2010qa,Dennen:2010dh,Huang:2011um,Plefka:2014fta}. This allowed one 
to use these powerful techniques to study tree-level scattering amplitudes in a concise fashion \cite{Huang:2011um,Plefka:2014fta}. 
The analysis in Ref.\ \cite{Plefka:2014fta} went further in multiplicity than any other study. In particular, making use of a numerical 
implementation of BCFW recursion relations \cite{Britto:2004ap,Britto:2005fq}, multiple forms of the six-leg amplitude were proposed there. 
Presently, we revisit its calculation by employing a different technique. Namely, constructing its ansatz in terms of dual conformally-covariant 
building blocks, we then fixed unknown accompanying coefficients by relying on its (assumed) universal double-collinear behavior, with 
corresponding splitting amplitudes deduced recently in Ref.\ \cite{Belitsky:2024rwv}, and 
further constraints from four-dimensional massless limits. Our finding differs from \cite{Plefka:2014fta}. Let us point out that the use of 
anticipated analytic properties of scattering amplitudes to constrain or fix their form is not new. It was used in the past to prove the form 
of next-to-maximal-helicity violating amplitudes on the conformal branch of the $\mathcal{N} = 4$ sYM \cite{Korchemsky:2009hm}. 
Further, taking the triple-collinear limit of the found six-leg amplitude, we calculated the triple-collinear splitting superamplitude.

Our subsequent presentation is organized as follows. First, we give the basics of the $\mathcal{N} = (1,1)$ on-shell superspace,
spinor-helicity formalism, and dual conformal properties of scattering amplitudes. We next, provide a detailed derivation of the
double-collinear splitting superamplitude in Sect.\ \ref{DoubleCollinearSection} starting from the five-leg amplitude. We assumed 
that it is universal for any number of legs and then used its form in Sect.\ \ref{SectionRefined} to find constraints on the six-leg 
amplitude. There is still some ambiguity left in the ansatz, so we projected out it on a component that involved only scalar external 
states. Since the latter depends only on Lorentz invariant six-dimensional products of particle momenta it can be obtained by an 
uplift of its four-dimensional counterpart. This projection then completely fixes the remaining freedom. Using so-obtained expressions 
as a starting point, we extracted from it a triple-collinear splitting amplitude. Finally, we provided an outlook for future use of our
results. Some appendices were added with details clarifying calculations in the main text.

\section{Symmetries of $\mathcal{N} = (1,1)$ superamplitudes}

The $\mathcal{N} = (1,1)$ sYM is a maximally supersymmetric gauge theory in six dimensions. All of its on-shell states are classified 
according to the little group $\mbox{SU}(2) \times \mbox{SU}(2)$ of its Lorentz group $\mbox{SO}(5,1) \simeq \mbox{SU}^\ast(4)$ 
\cite{Cheung:2009dc,Dennen:2009vk}. The $R$-symmetry of the model is $\mbox{SU}_{R} (2) \times \mbox{SU}_{R} (2)$, however, 
it would require additional indices for the enumeration of physical degrees of freedom and result in their proliferation far beyond the number 
of available states. Thus, one sacrifices the $R$-group in favor of the little group such that only its $\mbox{U}_{R} (1) \times \mbox{U}_{R} (1)$ 
subgroup is left manifest, i.e., one imposes a truncation. The on-shell states of the model
\begin{align*}
&\mbox{gluons:} &&  g^a{}_{\dot{a}} \, , \\
&\mbox{scalars:} &&  \phi, \phi^\prime, \phi^{\prime\prime}, \phi^{\prime\prime\prime}  \, , \\
&\mbox{gluinos:} &&  \chi^a, \bar\chi_{\dot{a}}, \psi^a, \bar\psi_{\dot{a}}  \, ,
\end{align*}
can be packaged into a single CPT self-conjugate non-chiral superfield \cite{Dennen:2009vk}
\begin{align}
\label{6Dsuperfield}
{\mit\Phi} = \phi + \chi^a \eta_a + \bar{\chi}_{\dot{a}} \bar\eta^{\dot{a}} + \eta^2 \phi^\prime + \bar\eta^2 \phi^{\prime\prime}
+
g^{a}{}_{\dot{a}} \eta_a \bar\eta^{\dot{a}} 
+ 
\psi^a \eta_a \bar\eta^2
+
\bar\psi_{\dot{a}} \bar\eta^{\dot{a}} \eta^2
+
\eta^2 \bar\eta^2 \phi^{\prime\prime\prime}
\, ,
\end{align}
as a terminating expansion in the complex, independent Grassmann variables $\eta_a$ and $\bar\eta_{\dot{a}}$ that carry only the 
little group indices and  possess positive and negative chirality
\begin{align}
\eta \to e^{i \phi} \eta
\, , \qquad
\bar\eta \to e^{-i \phi} \bar\eta
\, .
\end{align}

Scattering amplitudes in the theory are generated from the amputated vacuum expectation value of a product of superfields,
schematically
\begin{align}
\label{6Damplitude}
\mathcal{A}_n 
&  
=
\vev{{\mit\Phi}_1 \dots {\mit\Phi}_n} 
\, .
\end{align}
They depend on $n$ bosonic momenta $P_i$ and $n+n$ of (anti-)chiral charges $Q_i$ and $\bar{Q}_i$, cumulatively called 
supermomenta. The theory benefits from a spinor-helicity formalism \cite{Cheung:2009dc}, which allows one to recast the 
super-Poincar\'e quantum numbers in terms of unconstrained Weyl spinors $\Lambda_i^{A, a} \equiv \ket{i^{a}} = \bra{i^{a}}$ 
and $\bar\Lambda_{i, A, \dot{a}} \equiv |i_{\dot{a}}] = [i_{\dot{a}}|$,
\begin{align}
\label{SpinorHelicityRep}
P_i = \ket{i^{a}} \bra{i_a}
\, , \qquad
\bar{P}_i = |i_{\dot{a}}] [i^{\dot{a}}|
\, , \qquad
Q_i
=
\bra{i^{a}} \eta_{i, a}
\, , \qquad
\bar{Q}_i
=
[i_{\dot{a}}| \bar\eta_i^{\dot{a}}
\, .
\end{align}
Our conventions for these were thoroughly spelled out in Ref.\ \cite{Belitsky:2024rwv} and will not be repeated here. Imposing the 
super-momentum conservation, we can extract it in terms of bosonic and fermionic delta functions\begin{align}
\mathcal{A}_n 
=
i (2 \pi)^6 
\delta^{(6)} \left( \sum\nolimits_{i = 1}^n P_i \right) 
\delta^{(4)} \left( \sum\nolimits_{i = 1}^n Q_i \right)
\delta^{(4)} \left( \sum\nolimits_{i = 1}^n \bar{Q}_i \right)
\widehat{\mathcal{A}}_n 
\, ,
\end{align} 
and define reduced amplitudes $\widehat{\mathcal{A}}_n$ which are homogeneous polynomials of order $n - 4$ in Grassmann
variables. The non-chiral nature of the theory imposes a more stringent constraint on these: they can be chosen to be polynomials
of equal degrees $[n/2] - 2$ both in the chiral and anti-chiral charges \cite{Plefka:2014fta} yielding natural reduction properties to
four dimensions.

Since the six-dimensional Yang-Mills theory possesses dimensionful gauge coupling, it is not classically invariant under
dilatations and special conformal boosts. This translates into the same properties for tree-level scattering. However,
making use of the supersymmetric generalization of BCFW recursion relations, it was demonstrated in Refs.\ 
\cite{Bern:2010qa,Dennen:2010dh,Plefka:2014fta} that the reduced tree-level amplitudes $\widehat{\mathcal{A}}_n^{(0)}$ 
transform covariantly under the dual conformal inversion $\mathcal{I}$ with the same conformal weight for all legs,
\begin{align}
\label{ConformalInversionAhat}
\mathcal{I} \widehat{\mathcal A}^{(0)}_n = X_1^2 \dots X_n^2 \widehat{\mathcal A}^{(0)}_n
\, .
\end{align}
Here, $X_i$ are the region momenta, aka dual coordinates, defined along with their supersymmetric counterparts ${\mit\Theta}_i$
and $\bar{\mit\Theta}_i$ as
\begin{align}
\label{superPoincare2Dual}
P_i = X_{i i+1}
\, , \qquad
Q_i = {\mit\Theta}_{i i+1}
\, , \qquad
\bar{Q}_i = \bar{\mit\Theta}_{i i+1}
\, ,
\end{align}
with the adopted convention $\ast_{ij} \equiv \ast_i - \ast_j$ throughout this paper. The discrete inversion is defined on these as
\begin{align}
\label{Inversions}
\mathcal{I} X^{AB} = \frac{\bar{X}_{AB}}{X^2}
\, , \qquad
\mathcal{I} {\mit\Theta}^A = {\mit\Theta}^B \frac{\bar{X}_{BA}}{X^2} 
\, , \qquad
\mathcal{I} \bar{\mit\Theta}_A = \frac{X^{AB}}{X^2} \bar{\mit\Theta}_B 
\, ,
\end{align}
where we displayed the $\mbox{SU}^\ast (4)$ Lorentz indices explicitly for notational clarity. The inversion is an involution with
$\mathcal{I}^2 = 1$. Though the proof of Eq.\ \re{ConformalInversionAhat} via BCFW recursion relations requires, at intermediate 
stages, defining the operation of inversion on the spinor-helicity variables carrying the little group indices as well, we do not need 
to spell them out explicitly here since the reduced amplitudes ultimately depend only on super-Poincar\'e charges $P_i$, $Q_i$ 
and $\bar{Q}_i$ in light of their non-chiral nature. 

The fact that $\widehat{\mathcal A}_n$ are dual translationally invariant functions of the dual variables $(X_i, {\mit\Theta}_i,$ 
$\bar{\mit\Theta}_i)$ and covariant under the dual inversion provides severe constraints on their form. There is a unique way
to construct primary building blocks by uplifting a four-dimensional construction from Ref.\ \cite{Drummond:2008vq}.  
The required dual conformal covariants are \cite{Drummond:2008vq,Huang:2011um,Plefka:2014fta}
\begin{align}
\label{BconfCovs}
\bra{B_{i,jk}} = \bra{{\mit\Theta}_{ij}} \bar{X}_{jk} X_{ki} + \bra{{\mit\Theta}_{ik}} \bar{X}_{kj} X_{ji}
\, , \qquad
|\bar{B}_{i,jk}] = - {\rm cc} \big( \bra{B_{i,jk}} \big)
\, ,
\end{align}
where $\mbox{cc}$ stands for the chiral conjugate, i.e., changing all unbarred symbols with barred and vice versa. The (square) 
kets are defined as minus the chiral conjugate of (angle) bras, as in Ref.\ \cite{Plefka:2014fta}. These invert as
\begin{align}
\mathcal{I} \bra{B_{i,jk}} = \frac{\bra{B_{i,jk}} \bar{X}_i}{X_i^2 X_j^2 X_k^2}
\, ,
\end{align}
under Eqs.\ \re{Inversions}. Though one may sandwich an odd number of momenta between bras and kets of the same chirality
and define chiral conjugate pairs of conformal covariants; these lead to a decoupling of chiral and anti-chiral degrees of freedom 
and, therefore, can be eliminated from potential candidates of viable building blocks due to the non-chiral nature of the 
$\mathcal{N} = (1,1)$ sYM. Similarly, one could sandwich an even number of momenta between the brackets of opposite chirality 
as more generic ingredients, but the one with just a unit matrix in between them is a good, minimal starting point \cite{Plefka:2014fta},
\begin{align}
\label{MinimalElements}
\Omega_{ijklm} = \ft12  \bra{B_{i,jl}} \bar{B}_{i,km}] + {\rm cc}
\, .
\end{align}
The inner product involved acquires the same conformal weights for all indices. These objects are of Grassmann degree $1+1$.

Color-ordered tree amplitudes exhibit simple analytic behavior when the invariant mass formed by adjacent momenta goes on its 
mass shell: they develop poles in corresponding Mandelstam invariants $S_{i i+1} \equiv (P_i + P_{i+1})^2$. In conjunction with the known 
Grassmann degree of $n$-leg amplitudes and their dual inversion properties, one can predict the four-leg amplitude unambiguously 
(up to an overall phase) \cite{Cheung:2009dc,Dennen:2009vk,Plefka:2014fta} and put forward a compact ansatz for the five-leg case
\begin{align}
\label{CompactA5}
\widehat{\mathcal{A}}_4^{(0)} 
= \frac{1}{S_{12} S_{23}}
\, , \qquad
\widehat{\mathcal{A}}_5^{(0)} 
=
\frac{- \Omega_{12345}}{S_{12} S_{23} S_{34} S_{45} S_{51}}
\, .
\end{align} 
The latter was proposed and confirmed with the numerical implementation of BCFW recursion 
\cite{Plefka:2014fta}. This representation is to be contrasted with a lengthy multi-term expression devised in \cite{Dennen:2009vk}, and 
quoted in Eq.\ \re{5legTree6D} for readers' convenience. The latter does not exhibit obvious dual conformal properties, however, with a 
little work, the two can be analytically shown to be equivalent, see Appendix \ref{AppendixEquivalence}.

\section{Double-collinear splitting superamplitude}
\label{DoubleCollinearSection}

To prepare the ground for attacking the six-leg amplitude $\widehat{\mathcal{A}}_6^{(0)}$, we first extract 
the double-collinear
splitting amplitude from the five-leg one. We will implement this limit on legs 4 and 5. We expect that $\widehat{\mathcal{A}}_5^{(0)}$ 
admits a factorized form
\begin{align}
\widehat{\mathcal{A}}_5^{(0)} (P_1, P_2, P_3, P_4, P_5)
\xrightarrow{\scriptscriptstyle 4||5}
\widehat{\mathcal{A}}_4^{(0)} (P_1, P_2, P_3, P)
{\mathcal S}\mbox{plit}^{(0)} (- P; P_4, P_5)
\, ,
\end{align}
where the leg $P = P_4 + P_5$ in the super-splitting amplitude ${\mathcal S}\mbox{plit}$ is slightly off-shell, i.e., $P^2 \neq 0$. Thus, 
one of the Mandelstam invariants $S_{45} = P^2$ is `small' with the rest being `large'.

To study the above asymptotics in a self-consistent manner, we need to specify how the collinear configuration is approached. This can 
be done with a Sudakov parametrization of the involved momenta $P_4$ and $P_5$ as was done in Ref.\ \cite{Catani:2011st} in
four dimensions. Namely, these light-like momenta are decomposed as
\begin{align}
\label{CollMomSudakDecomp}
P_4 = z P + K_\perp - \frac{K_\perp^2}{2 z (P \cdot N)} N
\, , \qquad 
P_5 = \bar{z} P - K_\perp - \frac{K_\perp^2}{2 z (P \cdot N)} N
\, ,
\end{align}
where the two momenta $P_{4,5}$ approach a common light-like direction $P$, $P^2 = 0$, with corresponding momentum fractions
being $z$ and $\bar{z} \equiv 1 - z$, respectively. The vector $N$ is another light-cone momentum, $N^2 = 0$, moving in the `opposite' 
direction to $P$ and is required in order to define the transverse momentum $K_\perp$ that parametrizes how the common collinear 
direction is reached, $K_\perp \cdot P = K_\perp \cdot N = 0$. The inner product of the two momenta decays quadratically in $K_\perp \to 0$
\begin{align}
P_4 \cdot P_5 = - \frac{K_\perp^2}{2 z \bar{z}}
\, ,
\end{align}
so that $\langle 4_a | 5_{\dot{a}}] \sim \langle 5_a | 4_{\dot{a}}] \sim O (K_\perp/\sqrt{z \bar{z}})$. Since the amplitude
diverges as $1/S_{45} \sim 1/K_\perp^2$ in the collinear limit, the momentum algebra in the numerator of the amplitude
has to be performed to linear order in $K_\perp$ to extract the splitting amplitude. To this accuracy, we will use
\begin{align}
\label{MatrixCollLimit}
P_4 \bar{P}_5 = - P_5 \bar{P}_4 = K_\perp \bar{P} + O (K_\perp^2)
\, ,
\end{align}
for the matrix products.

With the above results in our hands, we start inspecting the five-leg amplitude in the form of Eq.\ \re{SimplifiedOmega}. It is expected, 
based on universal factorization properties of gauge amplitudes, that none of the fermionic structures encoding particles other than 
$4$ and $5$ contribute to the $4||5$ collinear limit. This is obvious from the representation \re{SimplifiedOmega} since the `alien' structures 
are accompanied by the vanishing Mandelstam invariant $S_{45} \to 0$, i.e., the first line in Eq.\ \re{SimplifiedOmega} provides a regular 
contribution to the amplitude and is not relevant.  Thus, we turn to the remaining ones next. Let us start with the diagonal terms, i.e., the 
first two terms in the second line. Using the relations $S_{51} + S_{41} + S_{45} = S_{23}$ and $S_{34} + S_{35} + S_{45} = S_{12}$ 
between the Mandelstam invariants, we can rewrite them as
\begin{align}
S_{51} Q_4 \bar{P}_3 P_5 \bar{Q}_4 
&
= 
S_{23} Q_4 \bar{P}_3 P_5 \bar{Q}_4 + Q_4 \bar{P}_1 K_\perp \bar{P} P_3 \bar{Q}_4
+
O (K_\perp^2)
\, , \\
S_{34} Q_5 \bar{P}_4 P_1 \bar{Q}_5
&
=
S_{12} Q_5 \bar{P}_4 P_1 \bar{Q}_5 + Q_5 \bar{P}_1 K_\perp \bar{P} P_3 \bar{Q}_5
+
O (K_\perp^2)
\, .
\end{align}
Notice that the first term in both equations is of order $K_\perp$ since they have momenta and chiral charges of coalescing
states adjacent to each other and they are of the same order as the second terms. All that is left to analyze in Eq.\ \re{SimplifiedOmega} 
is the last term in the second line. We find for it
\begin{align}
Q_4 [S_{34} S_{51} &- \bar{P}_3 P_5 \bar{P}_4 P_1] \bar{Q}_5
=
Q_4 [\bar{P}_3 P_4 \bar{P}_5 P_1 - \bar{P}_3 P_5 \bar{P}_4 P_1] \bar{Q}_5
\\\
&
=
2 Q_4 \bar{P}_3 K_\perp \bar{P} P_1 \bar{Q}_5 + O (K_\perp^2)
=
2 Q_4 \bar{P}_1 K_\perp \bar{P} P_3 \bar{Q}_5 + O (K_\perp^2)
\, . \nonumber
\end{align}
Adding our findings together, we conclude that the numerator of the five-leg amplitude reduces in the $4||5$ double-collinear limit to
\begin{align}
\label{Coll45Asy}
- 
\Omega_{51234}
\xrightarrow{\scriptscriptstyle 4||5}
- 
\Omega^{||}_{51234}
=
S_{23} Q_4 \bar{P}_3 P_5 \bar{Q}_4
&
+
S_{12} Q_5 \bar{P}_4 P_1 \bar{Q}_5
\\
&
+
( Q_4 + Q_5 )\bar{P}_1 P_4 \bar{P}_5 P_3 (\bar{Q}_4 + \bar{Q}_5)
+
O (K_\perp^2)
\, . \nonumber
\end{align}
Here in the second term, we restored the near-collinear momenta $P_{4,5}$ from $K_\perp$ and $P$ by using Eq.\ \re{MatrixCollLimit}
backwards, i.e., from right to left. 

Finally, we need to eliminate the `alien' momenta $P_1$ and $P_3$ from all Dirac strings. This is accomplished by 
using the spinor-helicity representation \re{SpinorHelicityRep} for the super-momenta and factoring out `small', i.e., $O(K_\perp)$, 
Weyl inner products first and then imposing the strict collinear kinematics, i.e., just the first term in Eq.\ \re{CollMomSudakDecomp}, 
on the rest of the accompanying kinematical factors. For instance, for the first term in Eq.\ \re{Coll45Asy} we get
\begin{align}
Q_4 \bar{P}_3 P_5 \bar{Q}_4 
\simeq 
\sqrt{z \bar{z}} \eta_4^a \bra{P_a} \bar{P}_3 \ket{P_b} \bra{5^b} 4_{\dot{a}}] \bar\eta_4^{\dot{a}}
\simeq
-
\sqrt{z \bar{z}} S_{12} \bra{5_a} 4_{\dot{a}}] \eta_4^a \bar\eta_4^{\dot{a}}
\, .
\end{align}
Here, after the first equality sign, we used the strict collinearity conditions at the level of the six-dimensional Weyl spinors (and their 
chiral conjugates)
\begin{align}
\bra{4^a} = \sqrt{z} \bra{P^a}
\, , \qquad
\bra{5^a} = \sqrt{\bar{z}} \bra{P^a}
\, .
\end{align}
While, after the second one, we employed the relation $\bra{P_a} \bar{P}_3 \ket{P_b} = \varepsilon_{ab} S_{12}$ with $S_{3P} = S_{12}$
stemming from the momentum conservation condition in the accompanying four-leg amplitude $P_1+P_2+P_3+P = 0$. Similarly, we 
proceed with the rest of the terms in Eq.\ \re{Coll45Asy}. In the last, one has to commute $P_{4,5}$ to the left/right to pull out a `large' 
Mandelstam invariant from the Dirac string, first. Notice that to the $O(K_\perp^2)$ accuracy, the two inner products $\langle 4_a | 5_{\dot{a}}] 
\simeq - \langle 5_a | 4_{\dot{a}}]$ are the same, up to a phase, as can be easily concluded from the six-dimensional Clifford algebra of the 
corresponding near-collinear momenta.

Summarising our findings and dividing out the reduced four-leg tree amplitude \re{CompactA5}, we deduce the double-collinear splitting 
superamplitude
\begin{align}
\label{TreeSuperSplittingAmplitude}
&
\mbox{$\mathcal{S}$plit}^{(0)} (- P; P_4, P_5)
=
\frac{- \Omega^{||}_{51234}}{z \bar{z}  S_{45} S_{12} S_{23}}
\\
&
\qquad\qquad
=
\frac{\langle 4_a | 5_{\dot{a}}]}{\sqrt{z \bar{z}} S_{45}}
\Big[
\eta_4^a \bar\eta_4^{\dot{a}}
+
\eta_5^a \bar\eta_5^{\dot{a}}
-
(\sqrt{z} \eta_4^a + \sqrt{\bar{z}} \eta_5^a) (\sqrt{z} \bar\eta_4^{\dot{a}} + \sqrt{\bar{z}} \bar\eta_5^{\dot{a}})
\Big]
\, , \nonumber
\end{align}
in agreement with Ref.\ \cite{Belitsky:2024rwv}.

\section{Six-leg superamplitude}

In this section, we are turning to one of the main objectives of this study: the construction of the six-leg superamplitude using
constraints from its anticipated behavior in the double-collinear limit. Namely, assuming the universality of the splitting 
amplitude ${\mathcal S}\mbox{plit}$ derived in the previous section, the equation
\begin{align}
\label{6To5Col}
\widehat{\mathcal{A}}_6^{(0)} (P_1, P_2, P_3, P_4, P_5, P_6)
\xrightarrow{\scriptscriptstyle 5||6}
\widehat{\mathcal{A}}_5^{(0)} (P_1, P_2, P_3, P_4, P)
{\mathcal S}\mbox{plit}^{(0)} (- P; P_5, P_6)
\, ,
\end{align}
imposes rather rigid constraints on the form of $\widehat{\mathcal{A}}_6^{(0)}$. Let us proceed in a step-wise fashion.

First, we have to come up with a suitable basis for its building blocks. At initialization, we can use the minimal elements 
\re{MinimalElements} for this purpose. This was the working hypothesis of Ref.\ \cite{Plefka:2014fta} as well. We chose 
the set of 6 inequivalent elements
\begin{align}
\label{SixElements}
\{ \Omega_1, \Omega_2, \Omega_3, \Omega_4, \Omega_5, \Omega_6 \}
\equiv
\{ \Omega_{12345}, \Omega_{23456}, \Omega_{34561}, \Omega_{45612}, \Omega_{56123}, \Omega_{61234} \}
\, ,
\end{align}
which we enumerated by the value of their header index. Since the six-leg reduced amplitude is of Grassmann degree $2+2$
so a natural ansatz for it, taking into account the anticipated pole structure in Mandelstam invariants of adjacent momenta, reads
\begin{align}
\widehat{\mathcal{A}}_6^{(0)} (P_1, P_2, P_3, P_4, P_5, P_6)
=
\frac{\mathcal{N}_6}{S_{12} S_{23} S_{34} S_{45} S_{56} S_{61}}
\end{align}
with
\begin{align}
\label{PreliminaryAnsatz}
\mathcal{N}_6
=
\sum\nolimits_{i,j = 1}^6 \alpha_{ij} \Omega_i \Omega_j
\, ,
\end{align}
Here, the $\alpha_{ij} $ coefficients are some functions of $S_{ij}$. It is our goal to constrain or fix their form.

\subsection{Double-collinear behavior}

We will postpone the discussion of dual inversion properties to later stages and start with the analysis of the $5||6$ double-collinear 
behavior of $\Omega_i$ that we will name $\Omega^{||}_i$. A consideration akin to the one performed in the previous section 
demonstrates that out of the six elements \re{SixElements}, half of them are `large' and the other half are `small'. Namely, the `large' 
ones are of order $O(K_\perp^0)$ in the collinear kinematics and reduce to the $\Omega$-structure of the five-leg amplitude 
\re{CompactA5} with label $5 = P$,
\begin{align}
\label{LargeOmegaRel}
\Omega_6^{||} = \bar{z}\, \Omega_{P1234} = \bar{z}\, \Omega
\, , \qquad
\Omega_1^{||} = \Omega_{1234P} = \Omega
\, , \qquad
\Omega_2^{||} = z \, \Omega_{234P1} = z \, \Omega
\, .
\end{align}
Here we exhibited the ordering of labels in the intermediate expressions before we ultimately used the cyclic symmetry
of the $\Omega$-covariants \re{CyclicOmega5} in the five-leg amplitude. The `small' ones, $\Omega^{||}_{3,4,5}$ are of 
order $O(K_\perp^1)$. A linear superposition of $\Omega^{||}_{3}$ and $\Omega^{||}_{5}$ is related to $\Omega^{||}_{4}$ 
via the equation
\begin{align}
\label{SmallOmegaRel}
S_{61} \Omega^{||}_{3}
+
S_{45} \Omega^{||}_{5}
=
S_{345}
\Omega^{||}_{4}
\, , 
\end{align}
where we introduced a three-particle Mandelstam invariant which can be related to the distance in dual coordinates as $S_{345} 
= X_{36}^2$. $\Omega_4^{||}$ generates the numerator structure of the splitting amplitude \re{Coll45Asy},
\begin{align}
{\mathcal S}\mbox{plit}^{(0)} (-P; P_5, P_6)
=
\frac{- \Omega_4^{||}}{z \bar{z} S_{56} S_{4P} S_{P1}} 
\, ,
\end{align}
as in Eq.\ \re{TreeSuperSplittingAmplitude}.

Collinear limits in the other five nearest-neighbor channels can be found from the above by a cyclic permutation.

\subsection{Refined ansatz}
\label{SectionRefined}

Having established the collinear behavior of $\Omega$'s in one particular channel is sufficient to `weed out' redundant
contributions in the previous crude ansatz \re{PreliminaryAnsatz}. The factorized form of \re{6To5Col} suggests that
the numerator of the six-leg amplitude when legs 5 and 6 coalesce can have products of `large'$\times$`small' and
`small'$\times$`small' $\Omega$'s but not `large'$\times$`large' as this would lead to way too strong collinear singularity, 
unless the latter is accompanied by a `small' Mandelstam invariant\footnote{There is a remote possibility that some linear 
combination of `large' $\Omega$'s yields a `small' one as an $O(K_\perp^1)$ effect, but an inspection of higher terms in 
their expansion suggests that this is highly unlikely and thus can be disregarded in the first attempt to construct the 
numerator $\mathcal{N}_6$.} in the amplitude. Thus, we choose a `large' $\Omega_i$ with the largest index 
$i (\mbox{mod}\, 6)$ and multiply it by a linear combination of the `small' ones. This yields
\begin{align}
\label{RefinedAnsatz}
\mathcal{N}_6 = \Omega_2 \left( \alpha_1 \Omega_3 + \alpha_2 \Omega_4 + \alpha_3 \Omega_5 \right)
+ \mbox{cyclic permutations}
\, ,
\end{align}
with unknown functions $\alpha_i$ of the Mandelstam variables. In this manner, out of 36 terms in Eq.\ \re{PreliminaryAnsatz} 
only 18 survive, and there are only three truly unknown coefficients. Of course, terms obtained by cyclic permutations do
contain the products of two `large' $\Omega$'s and seem to invalidate this form if corresponding $\alpha$'s are `large'. This
does not happen, however, as we demonstrate next.

To constrain the form of $\alpha$'s, we turn next to the dual covariant properties of the amplitude \re{ConformalInversionAhat}.
Let us inspect the dual inversion of the products of $\Omega$'s entering Eq.\ \re{RefinedAnsatz}. In $\Omega_2 
\Omega_3$, $\Omega_2 \Omega_4$, and $\Omega_2 \Omega_5$, there are two labels that are not common to both and, thus, 
lead to mismatching powers of dual coordinates upon inversion. These arise for the labels $2\, \&1$, $3\, \&1$ and $4\, \&1$, 
respectively. To have the same weight for all contributing structures, we need to `extract' proper combinations of dual distances 
from $\alpha$'s. Notice, however, that the conformal weights of the two-particle Mandelstam invariants would compensate completely the ones 
from the thus-constructed numerator. So where will the required inversion weight on the right-hand side of Eq.\ \re{ConformalInversionAhat} 
come from? It can only stem from the three-particle invariants, i.e., $S_{123} = S_{456} = X_{14}^2$, $S_{234} =S_{561} = X_{25}^2$ 
and $S_{345} = S_{612} = X_{36}^2$, introduced as a product in the denominator of the reduced amplitude. This does not violate the 
two-particle factorization of the amplitude. However, it brings in three-particle poles into the game. These can and do arise in six-leg 
amplitudes but they cannot simultaneously happen in partially overlapping channels in light of Steinmann relations\footnote{At the 
tree level we are working with, this is obvious since there can be only one pole from $3\to1$ transition in a given Feynman graph.}
\cite{Steinmann,Caron-Huot:2016owq}, which imposes rigid constraints on the structure of the numerator $\mathcal{N}_6$. Having this
in mind, we rescale the $\alpha_i$ in Eq.\ \re{RefinedAnsatz} as
\begin{align}
\label{RescaledAlphas}
\alpha_i \to \frac{\alpha_i}{X_{14}^2 X_{25}^2 X_{36}^2}
\, .
\end{align}

Now, we fulfill the goal of recovering the same conformal weight for individual terms in \re{RefinedAnsatz}.
Starting with $\Omega_2 \Omega_3$, we notice that an obvious choice of a single power of $X_{12}^2$ is a no-starter since 
this distance is light-like. Thus, the labels $1$ and $2$ should belong to two different squared distances $X_{1i}^2 X_{2j}^2/X_{ij}^2$ 
with $i$ and $j$ compensated by an appropriate denominator $X_{ij}^2$. Now, from the singularity structure of the double collinear 
limit, we know that we cannot have $i$ and $j$ to be next-to-nearest neighbors as this would result in a double pole in the 
corresponding Mandelstam variable forbidden by Eq.\ \re{6To5Col}. Thus, the only option is to have $|i - j| = 3$, which together
with the fact that $i$ and $j$ cannot be adjacent to $1$ and $2$ uniquely fixes them to $i = 3$ and $j = 6$. We conclude
\begin{align}
\label{Alpha1}
\alpha_1 = \frac{X_{13}^2 X_{26}^2}{X_{36}^2} a_1
\, ,
\end{align}
where $a_1$ is not a constant, rather, it is a possible function of dual conformal cross-ratios that one can introduce starting from 
six legs. These are
\begin{align}
\label{CrossRatios}
U_1 = \frac{X_{13}^2 X_{46}^2}{X_{14}^2 X_{36}^2}
\, , \qquad
U_2 = \frac{X_{24}^2 X_{15}^2}{X_{25}^2 X_{14}^2}
\, , \qquad
U_3 = \frac{X_{35}^2 X_{26}^2}{X_{36}^2 X_{25}^2}
\, .
\end{align}
Notice that they transform into each other under the cyclic permutation $i \to i+1$,
\begin{align}
U_i \to  U_{i+1 ({\rm\scriptstyle mod}\, 3)}
\, .
\end{align}

Next, we turn to $\Omega_2 \Omega_3$ where $3\, \&1$ are not common so that we restore the same conformal weight for
all labels with
\begin{align}
\label{Alpha2}
\alpha_2 = X_{13}^2 a_2
\, ,
\end{align}
with $a_2$ being potentially a function of $U_i$.

Finally, we consider $\Omega_2 \Omega_3$ where the mismatched labels $4$ and $1$ are far-distant in a six-leg amplitude, 
therefore,
\begin{align}
\label{Alpha31}
\alpha_3 = X_{14}^2 a_3
\, .
\end{align}
Now, considering the $5||6$ collinear limit, we observe that to have any chance of getting the double-collinear amplitude from
the linear combination of $\Omega^{||}_3$ and $\Omega^{||}_5$, $a_3$ should be a linear function of the conformal cross 
ratio $U_1$ since it contains the proper Mandelstam invariant $S_{45}$ that accompanies $\Omega_5^{||}$ in Eq.\ \re{SmallOmegaRel}.
Linear dependence on $U_3$ is ruled out because it is a `large' variable not possessing proper dependence on two-particle invariants. 
However, $U_2$ is not as it vanishes linearly as a power of $S_{56}$ and is thus a subleading $O(K_\perp^2)$ effect in the $5||6$ kinematics. 
Therefore, we further decompose $a_3$ as
\begin{align}
\label{Alpha32}
a_3 = U_1 a_{3,1} + U_2 a_{3,2}
\, ,
\end{align}
where $a_{3,i}$ can depend on the cross ratios \re{CrossRatios}.

Substitution of the constraint equations \re{RescaledAlphas}, \re{Alpha1}, \re{Alpha2}, \re{Alpha31} and \re{Alpha32} into Eq.\ 
\re{RefinedAnsatz} is the first stage of what we dub as the {\sl collinear bootstrap}. An inspection of the cyclic permutations of the 
first line in \re{RefinedAnsatz} immediately demonstrates that all contributions from products of two `large' $\Omega$'s are 
necessarily accompanied by the vanishing Mandelstam invariant $S_{56}$. Thus the dual-covariance properties of the amplitude 
ruled out a potential predicament in the ansatz from such terms. Finally, since the extracted denominator in Eq.\ \re{RescaledAlphas}
is symmetric under $i \to i+1 ({\rm mod}\, 3)$, it is only natural to assume that $a = \{a_1, a_2, a_{3,i}\}$ are symmetric functions
of the conformal cross ratios \re{CrossRatios},
\begin{align}
a = a (U_{\sigma(1)}, U_{\sigma(2)}, U_{\sigma(3)}) 
\, ,
\end{align}
with $\sigma \in S_3$.

Now, we are in a position to carefully match our ansatz in the $5||6$ limit to the right-hand side of Eq.\ \re{6To5Col}. Let us first collect
all terms proportional to the `large' $\Omega_2^{||}$. Stripping the latter along with the denominator in Eq.\ \re{RescaledAlphas} 
we find for them
\begin{align}
\frac{X_{13}^2}{X_{36}^2} \left[ a_1 X_{26}^2 \Omega_3^{||} + 2 a_{3,1} X_{46}^2 \Omega_5^{||} \right] + X_{13}^2 a_2 \Omega_4^{||}
\, .
\end{align}
Comparing the square bracket to Eq.\ \re{SmallOmegaRel}, we conclude that 
\begin{align}
\label{a31}
a_{3,1} = \ft12 a_1
\, ,
\end{align}
since then, the numerator of the double-splitting amplitude factorizes out producing
\begin{align}
X_{13}^2 (a_1 + a_2) \Omega_4^{||}
\, .
\end{align}
Similarly, we analyze the collinear behavior of the terms involving `large' $\Omega^{||}_1$ and $\Omega^{||}_6$. To form and factor
out the splitting amplitude, we need to impose a condition similar to \re{a31} on $a_{3,2}$ as well,
\begin{align}
\label{a32}
a_{3,2} = \ft12 a_1
\, .
\end{align}
Adding everything together we obtain
\begin{align}
\mathcal{N}_6^{||}
=
\frac{ \Omega_4^{||}}{X_{14}^2 X_{25}^2 X_{36}^2}
\left[
(a_1 + a_2)
X_{13}^2 \Omega_2^{||}
+
(a_1 + a_2)
X_{35}^2 \Omega_6^{||}
+
X_{36}^2 [a_2 + a_1 (U_1 + U_3)] \Omega_1^{||}
\right]
\, .
\end{align}
Finally, recalling that $X_{36}^2 \xrightarrow{\scriptscriptstyle 5||6} z X_{13}^2 + \bar{z} X_{35}^2$ along with $U_1 + U_3 
\xrightarrow{\scriptscriptstyle 5||6} 1$ in the collinear limit, we find
\begin{align}
\label{56onN6}
\mathcal{N}_6^{||}
=
\frac{\Omega_1^{||} \Omega_4^{||}}{S_{4P} S_{P1}}
\, ,
\end{align}
making use of Eq.\ \re{LargeOmegaRel} and the collinear constraint on the sum of the remaining two unknowns
\begin{align}
\label{a1a2coll}
2 (a_1 + a_2) \xrightarrow{\scriptscriptstyle 5||6} 1
\, .
\end{align}
Equation \re{56onN6} is of the form of the right-hand side of Eq.\ \re{6To5Col}. Our double-collinear bootstrap is unable, however, to 
fix the functions $a_1$ and $a_2$ individually.

\subsection{Scalar projection}

To alleviate the complication of the collinear bootstrap, we will fix the remaining freedom with a projection on a known component.
Upon dimensional reduction down to four space-time dimensions, \re{6Damplitude} generates all massless amplitudes when
the extra-dimensional components of particles' momenta are set to zero. It produces the maximal helicity-violating scattering as well
as all the rest, i.e., the non-maximal helicity-violating amplitudes. In this limit, $g^{2}{}_{\dot{1}}$ and $g^{1}{}_{\dot{2}}$ correspond to the 
positive and negative helicity gluons, while $g^{1}{}_{\dot{1}}$ and $g^{2}{}_{\dot{2}}$ are the four-dimensional scalars, which along 
with $\phi$, $\phi^\prime$, $\phi^{\prime\prime}$ and $\phi^{\prime\prime\prime}$ restore the $\mbox{SO}(6) \simeq \mbox{SU}(4)$ 
internal symmetry of the spinless sector. Without loss of generality, it is convenient to identify $\phi$ and $\phi^{\prime\prime\prime}$ 
with their four-dimensional counterparts $\phi^{34}$ and $\phi^{12}$ of the sextet $\phi^{AB}$.

Since the scattering amplitude involving only scalars depends solely on Mandelstam invariants, knowing its massless form in four 
dimensions will easily allow us to find it in six by an elementary uplift
\begin{align}
s_{ij \dots k} \to S_{ij \dots k}
\, ,
\end{align}
where $s_{ij \dots k}$ is a multiparticle Lorentz invariant in four dimensions. Thus, we consider
\begin{align}
\mathcal{A}^{4\rm D}_6
&  
=
\vev{\phi^{12}_1 \phi^{12}_2 \phi^{34}_3 \phi^{34}_4 \phi^{34}_5 \phi^{12}_6} 
\nonumber\\
&
=
\frac{\vev{45} [12] \bra{6} p_4 + p_5 |3]}{s_{456} \vev{56} [23] \bra{4} p_5 + p_6 |1]}
-
\frac{\vev{34} [61] \bra{2} p_3 + p_4 |5]}{s_{561} \vev{23} [56] \bra{4} p_5 + p_6 |1]}
\, ,
\end{align}
which is a next-to-maximal helicity-violating amplitude that was extracted from, e.g., Ref.\ \cite{Drummond:2008vq} in the 
spinor-helicity formalism. It can be converted to a form involving only Mandelstam invariants by multiplying the numerator 
and denominator with appropriate angle/square brackets and forming strings of Dirac matrices from these. In this manner, we get
\begin{align}
\label{4DscalarComponent}
\mathcal{A}^{4\rm D}_6  
=
\frac{1}{s_{56} s_{23}}
\left(
\frac{s_{12} s_{45}}{s_{456}} + \frac{s_{16} s_{34}}{s_{561}} - s_{612} 
\right)
=
\frac{x_{36}^2}{x_{15}^2 x_{24}^2} \left( u_1 + u_3 - 1 \right)
\, ,
\end{align}
where we displayed after the first equality sign the form that exhibits the singularity structure of the amplitude when groups of 
adjacent momenta become collinear, while after the second one, we showed it in terms of four-dimensional analogues of 
conformal cross ratios.

To extract this component from our ansatz \re{RefinedAnsatz}, we set the supercharges corresponding to legs 3, 4 and 5 to
zero,
\begin{align}
\label{Q345zero} 
\{ Q_3, Q_4, Q_5, \bar{Q}_3, \bar{Q}_4, \bar{Q}_5 \} = 0 
\, , 
\end{align}
solve $Q_1$, $Q_2$ as well as their chiral conjugates from the supermomentum-conserving delta functions in terms
of $Q_6$
\begin{align}
\label{Q12solQ6}
Q_1 = - \frac{Q_6 \bar{P_2} P_1}{S_{12}}
\, , \qquad
Q_2 = - \frac{Q_6 \bar{P_1} P_2}{S_{12}}
\, .
\end{align}
We list the thus-reduced non-vanishing $\Omega$'s, which we label $\Omega^{\rm sc}_i$, in Appendix \ref{AppendixScalarOmegas}. $Q_6$ 
and $\bar{Q}_6$ have now to be integrated out to project on the component in question. Introducing the short-hand notation
\begin{align}
\vev{\vev{ \dots}}_6 \equiv \int d^2 \eta_6 d^2 \bar\eta_6 \dots
\, ,
\end{align} 
the only relation one has to use is
\begin{align}
\label{Q6integrationOut}
\vev{\vev{Q_6^A Q_6^B \bar{Q}_{6, C} \bar{Q}_{6, D}}}_6  = P_6^{AB} \bar{P}_{6, CD}
\, ,
\end{align}
in the product of two $\Omega$'s. This converts the latter into lengthy Dirac traces which can be evaluated with, for example, 
{\tt FeynCalc} \cite{Mertig:1990an,Shtabovenko:2023idz} and yield
\begin{align}
\frac{
\vev{\vev{\Omega_1^{\rm sc} \Omega_4^{\rm sc}}}_6}
{X_{14}^3 X_{25}^2 X_{36}^2}
&
= \frac{X_{26}^2 X_{35}^2 X_{46}^2}{X_{13}^2} (1 - U_1 + U_2 - U_3)
\, , 
\\
\frac{
\vev{\vev{\Omega_1^{\rm sc} \Omega_5^{\rm sc}}}_6}
{X_{14}^3 X_{25}^2 X_{36}^2}
&
= \frac{X_{26}^2 X_{35}^2 X_{36}^2}{X_{13}^2} (1 - U_1 + U_2 - U_3)
\, , 
\\
\frac{
\vev{\vev{\Omega_1^{\rm sc} \Omega_6^{\rm sc}}}_6}
{X_{14}^3 X_{25}^2 X_{36}^2}&
= 2 \frac{X_{24}^2 X_{26}^2 X_{35}^2 X_{36}^2}{X_{13}^2 X_{25}^2} (1 - U_1)
\, , 
\\
\frac{
\vev{\vev{\Omega_4^{\rm sc} \Omega_5^{\rm sc}}}_6}
{X_{14}^3 X_{25}^2 X_{36}^2}
&
= 2 \frac{X_{15}^2 X_{26}^2 X_{36}^2 X_{46}^2}{X_{13}^2 X_{14}^2} (1 - U_3)
\, , 
\\
\frac{
\vev{\vev{\Omega_4^{\rm sc} \Omega_6^{\rm sc}}}_6}
{X_{14}^3 X_{25}^2 X_{36}^2}
&
= \frac{X_{26}^2 X_{36}^2 X_{46}^2}{X_{13}^2} (1 - U_1 + U_2 - U_3)
\, , 
\\
\frac{
\vev{\vev{\Omega_5^{\rm sc} \Omega_6^{\rm sc}}}_6}
{X_{14}^3 X_{25}^2 X_{36}^2}
&
= \frac{X_{26}^2 X_{36}^4}{X_{13}^2} (1 - U_1 + U_2 - U_3)
\, .
\end{align}
Substituting these expressions into our ansatz and equating it to the uplifted form of the six-scalar amplitude \re{4DscalarComponent}, 
we find that the only solution for the last two unknown functions $a_1$ and $a_2$ is
\begin{align}
a_1 = - \ft12 a_2 = \frac{1}{U_1 + U_2 + U_3 - 3}
\, .
\end{align}
These correctly reproduce the expected collinear behavior \re{a1a2coll}. We notice, however, that the six-leg amplitude develops a spurious 
pole at $U_1 + U_2 + U_3 = 3$, which should cancel upon extraction of individual components.

\subsection{Final expression}

Having completely fixed all unknown functions accompanying the dual covariants $\Omega_i$, we can finally present the six-leg amplitude.
It is
\begin{align}
\widehat{\mathcal{A}}_6^{(0)}
&
=
\frac{\mathcal{N}_6}{X_{13}^2 X_{24}^2 X_{35}^2 X_{46}^2 X_{51}^2 X_{62}^2 X_{14}^2 X_{25}^2 X_{36}^2 (U_1 + U_2 + U_3 - 3)} 
\, , 
\end{align}
with
\begin{align}
\label{N6numeratorFinal}
\mathcal{N}_6
&
=
\frac{X_{13}^2 X_{26}^2}{X_{36}^2} \Omega_2 \Omega_3 - 2 X_{13}^2 \Omega_2 \Omega_4 + \ft12 X_{14}^2 (U_1 + U_2) \Omega_2 \Omega_5
+
\mbox{cyclic permutations}
\, .
\end{align}
This expression disagrees with the one advocated in Ref.\ \cite{Plefka:2014fta} from the numerical implementation of six-dimensional BCFW 
recursion in the coefficient of the last term in the numerator, it is half of the one in that work. We also inspected another form of the alleged six-leg 
amplitude, given by Eq.\ (3.78) there. As it stands, it possesses the correct double-collinear behavior in the $5||6$ channel. However, it does not
yield the expected form when projected on the scalar component \re{4DscalarComponent}, which we extracted in the previous section. Namely, the 
term $\Omega_4 \Omega_5$ in Eq.\ (3.78) is an obstruction for an overall factorization of the spurious zero from the $(U_1 + U_3 - 1)$ factor in 
\re{4DscalarComponent}, provided one changes the relative sign in the factor accompanying the $\Omega_3 \Omega_6$ structure. Moreover, 
simply neglecting this term changes the location of the spurious pole to $U_1 - U_2 + U_3 = 1$, rather than $U_2 + U_3 - U_1 = 1$ reported there. 
These naive changes alter the proper collinear limit, however. It is unclear if one can salvage that formula.

\section{Triple-collinear splitting superamplitude}

Having constructed the six-leg amplitude, we can now consider its triple-collinear limit and extract the corresponding triple-collinear splitting 
superamplitude
\begin{align}
\label{6To4Col}
\widehat{\mathcal{A}}_6^{(0)} (P_1, P_2, P_3, P_4, P_5, P_6)
\xrightarrow{\scriptscriptstyle 4||5||6}
\widehat{\mathcal{A}}_4^{(0)} (P_1, P_2, P_3, P)
{\mathcal S}\mbox{plit}^{(0)} (- P; P_4, P_5, P_6)
\, .
\end{align}
As in Sect.\ \ref{DoubleCollinearSection}, we introduce the Sudakov decomposition for the near-collinear momenta
\begin{align}
P_i = z_i P + K_{i,\perp} - \frac{K_{i, \perp}^2}{2 z_i (P \cdot N)} N
\, , \qquad 
i = 4,5,6
\end{align}
with
\begin{align}
\sum_i z_i = 1
\, , \qquad \sum_i K_{i, \perp} = 0
\, ,
\end{align}
to have a clear identification of the $K_\perp$-scaling for various terms in this kinematics.

Since the reduced four-leg amplitude is just an overall scalar factor \re{CompactA5}, we strip it down and define the triple-splitting superamplitude 
as
\begin{align}
\label{TripleColSplittingSuperAmp}
{\mathcal S}\mbox{plit}^{(0)} (- P; P_4, P_5, P_6)
=
\frac{\mathcal{N}_6^{|||}/(S_{12} S_{23})^2}{z_4 z_6 S_{45} S_{56} [\bar{z}_4 S_{45} + \bar{z}_6 S_{56} + (z_4 z_6 - 3 \bar{z}_4 \bar{z}_6) S_{456}]}
\, .
\end{align}
The unfortunate feature of this representation is that the splitting superamplitude inherits the spurious pole of the six-leg amplitude.
Accepting this feature as unavoidable, it suffices to analyze the triple-collinear behavior of individual dual covariants $\Omega_i$
that we denote as $\Omega_i^{|||}$. Since the triple splitting has to have $1/S$-type singularity in the small Mandelstam invariants 
$S = \{  S_{45}, S_{56}, S_{456} \}$, the products $\Omega_i^{|||} \Omega_j^{|||}$ should scale at least as $O (K_\perp^2)$. The large 
kinematical denominator $(S_{12} S_{23})^2$ then has to cancel out with the same factors emerging from $\Omega_i^{|||}$. Last but not 
least, $\Omega_i^{|||}$ have to depend only on the (anti)chiral changes $Q_i/\bar{Q}_i$ with $i = 4,5,6$ and be oblivious to the rest.

Analyzing the scaling of $\Omega_i^{|||}$ with $K_\perp$, we find that the six dual covariants form two groups, one containing `small'
$\Omega_i^{|||} \sim O (K_\perp)$ with $i = 1,2,5,6$ and another one with two `ultra-small' ones $\Omega_i^{|||} \sim O (K_\perp^3)$ for 
$i = 3,4$. Dropping terms of orders higher than corresponding leading terms, the independence of $\Omega_i^{|||}$ on the (anti)chiral 
charges $Q_i/\bar{Q}_i$ with $i = 1,2,3$ becomes evident. In this manner, as an intermediate result, we find their leading asymptotics as $4||5||6$,
\begin{align*}
\Omega_1^{|||} 
&
= \ft12 (Q_4 + Q_5 + Q_6) (\bar{P}_4 + \bar{P}_5 + \bar{P}_6) P_1\left[ (\bar{P}_5 + \bar{P}_6) P_3 \bar{Q}_4 - S_{34} (\bar{Q}_5 + \bar{Q}_6) \right]
+ \mbox{cc}
\, , \\
\Omega_2^{|||} 
&
= \ft12
\left[
S_{23} (Q_5 + Q_6) - S_{234} (Q_4 + Q_5 + Q_6)
\right]
\bar{P}_3 (P_4 + P_5) (\bar{Q}_4 + \bar{Q}_5) 
+ \mbox{cc}
\, , \\
\Omega_3^{|||} 
&
= \ft12
(Q_4 + Q_5) (\bar{P}_4 + \bar{P}_5) P_3 
\left[
S_{56}
\bar{Q}_4
-
\bar{P}_4 (P_5 + P_6) (\bar{Q}_5 + \bar{Q}_6)
\right]
+ \mbox{cc}
\, , \\
\Omega_4^{|||} 
&
= \ft12
\left[
S_{56}
Q_4
-
(Q_5 + Q_6) (\bar{P}_5 + \bar{P}_6) P_4 
\right]
\left[
\bar{P}_5 P_1 \bar{Q}_6
-
S_{16} \bar{Q}_5
\right]
+ \mbox{cc}
\, , \\
\Omega_5^{|||} 
&
= \ft12
\left[
Q_6 \bar{P}_1 P_5
-
S_{16} Q_5
\right]
\left[
S_{12} \bar{Q}_6
-
\bar{P}_6 P_3
\left( \bar{Q}_4 + \bar{Q}_5 + \bar{Q}_5 \right)
\right]
+ \mbox{cc}
\, , \\
\Omega_6^{|||} 
&
= \ft12
\left[
S_{12} Q_6 
-
(Q_4 + Q_5 + Q_6) \bar{P}_3 P_6
\right]
\bar{P}_1 (P_4 + P_5 + P_6) \left( \bar{Q}_4 + \bar{Q}_5 + \bar{Q}_5 \right)
+ \mbox{cc}
\, .
\end{align*}
In this form, $\Omega$'s superficially possess residual dependence on the `alien' momentum labels $i = 1,2,3$. It is eliminated upon extraction
of `small' contractions $\bra{i}j]$, $i,j = 4,5,6$, and further imposing the strict triple-collinear limit in all factors accompanying them, i.e., 
setting $P_i = z_i P$. In this manner, we deduce the final expressions for $\Omega^{|||}$'s,
\begin{align}
\label{Omega1triple}
\Omega_1^{|||} 
&
= 
\ft12 S_{12} S_{23} \big[
\bra{4_a} 5_{\dot{a}}]  (\sqrt{z_4} \eta_5^a - \sqrt{z_5} \eta_4^a)
+
\bra{4_a} 6_{\dot{a}}]  (\sqrt{z_4} \eta_6^a - \sqrt{z_6} \eta_4^a )
+
\bra{5_a} 6_{\dot{a}}]  (\sqrt{z_5} \eta_6^a - \sqrt{z_6} \eta_5^a )
\big]
\nonumber\\
&
\times
\big[ \bar{z}_4 \sqrt{z_4} \eta_4^a - z_4 \sqrt{z_5} \eta_5^a - z_4 \sqrt{z_6} \eta_6^a \big]
+ \mbox{cc}
\, , \\
\Omega_2^{|||} 
&
= 
\ft12 S_{12} S_{23} 
\big[
\bar{z}_4 \sqrt{z_4} \eta_4^a
-
z_4 \sqrt{z_5} \eta_5^a
-
z_4 \sqrt{z_6} \eta_6^a
\big]
\bra{4_a}5_{\dot{a}}]
(\sqrt{z_4} \bar\eta_5^{\dot{a}} - \sqrt{z_5} \bar\eta_4^{\dot{a}})
+ \mbox{cc}
\, , \\
\Omega_3^{|||} 
&
= 
\ft12 S_{12} (z_4 \eta_5^a - \sqrt{z_4 z_5} \eta_4^a) \bra{4_a} 5_{\dot{a}}]
\big[ S_{56} \bar{\eta}_4^{\dot{a}} + (\bar{Q}_5 P_6 + \bar{Q}_6 P_5) |4^{\dot{a}}]  \big]
+ \mbox{cc}
\, , \\
\Omega_4^{|||} 
&
= 
\ft12 S_{23} \big[ S_{56} \eta_4^a + (Q_5 \bar{P}_6 + Q_6 \bar{P}_5) \ket{4^a} \big]
\bra{4_a} 5_{\dot{a}}]
(z_6 \bar{\eta}_5^{\dot{a}} - \sqrt{z_5 z_6} \bar{\eta}_6^{\dot{a}})
+ \mbox{cc}
\, , \\
\Omega_5^{|||} 
&
= 
\ft12 S_{12} S_{23} (\sqrt{z_5} \eta_6^a - \sqrt{z_6} \eta_5^a)
\bra{5_a} 6_{\dot{a}}]
\big[
z_6 \sqrt{z_4} \bar{\eta}_4^{\dot{a}}
+
z_6 \sqrt{z_5} \bar{\eta}_5^{\dot{a}}
-
\bar{z}_6 \sqrt{z_6} \bar{\eta}_6^{\dot{a}}
\big]
+ \mbox{cc}
\, , \\
\label{Omega6triple}
\Omega_6^{|||} 
&
= 
\ft12 S_{12} S_{23} \big[ z_6 \sqrt{z_4} \eta_4^a + z_6 \sqrt{z_5} \eta_5^a - \bar{z}_6 \sqrt{z_6} \eta_6^a\big]
\\
&
\times
\big[
\bra{4_a} 5_{\dot{a}}]  ( \sqrt{z_4} \bar{\eta}_5^{\dot{a}} - \sqrt{z_5} \bar{\eta}_4^{\dot{a}} )
+
\bra{4_a} 6_{\dot{a}}]  ( \sqrt{z_4} \bar{\eta}_6^{\dot{a}} - \sqrt{z_6} \bar{\eta}_4^{\dot{a}} )
+
\bra{5_a} 6_{\dot{a}}]  ( \sqrt{z_5} \bar{\eta}_6^{\dot{a}} - \sqrt{z_6} \bar{\eta}_5^{\dot{a}} )
\big]
+ \mbox{cc}
\, . \nonumber
\end{align}
Their substitution into Eq.\ \re{N6numeratorFinal} with the simultaneous replacement of all dual distances with their strict collinear limits
(except for the vanishing invariants),
\begin{align}
\label{TripleMandelstamInv}
&
\{X_{13}^2, X_{24}^2, X_{35}^2, X_{46}^2, X_{51}^2, X_{62}^2, X_{14}^2, X_{25}^2, X_{36}^2\}
\\
&
\xrightarrow{\scriptscriptstyle 4||5||6}
\{ S_{12}, S_{23}, z_4 S_{12}, S_{45}, S_{56}, z_6 S_{23}, S_{456} , \bar{z}_4 S_{23} , \bar{z}_6 S_{12} \}
\, , \nonumber
\end{align}
yields the final expression for the triple-splitting amplitude. Since this expression is merely a sum of its parts and is extremely lengthy,
we do not present it here to save space. As an ultimate check on these expressions, we considered the projection of 
${\mathcal S}\mbox{plit}^{(0)} (- P; P_4, P_5, P_6)$ on the purely scalar sector to observe the required cancellation of the spurious 
pole and confirm agreement with the expected form of the amplitude \re{4DscalarComponent}. This is demonstrated in 
Appendix~\ref{AppendixCollinearCheck}.

\section{Conclusions} 

In this paper, we studied six-dimensional amplitudes in $\mathcal{N} = (1,1)$ sYM theory in the multicollinear limits. We first used
known double-collinear splitting amplitudes to constrain the form of the six-leg case, which was previously extracted from
the numerical BCFW recursion relations. The collinear bootstrap, while strongly constraining its form, leaves some freedom in the
ansatz. The latter was fixed by considering a simple component of this superamplitude. Relying on this result, we extracted from it 
the triple-collinear splitting superamplitude.

The result of this consideration is of interest for multiple reasons. First, the use of the double- and triple-collinear behavior will 
undoubtedly help to construct the seven-leg amplitude in this theory and then find the quadruple-splitting amplitude from it. The 
process can then be repeated for higher multiplicity. The use of the known analyticity structure will certainly benefit from numerical 
BCFW recursions for the amplitude determination and these should be used in tandem.

The triple-splitting amplitude can now be used for the construction of integrands of the double-splitting amplitude. Previously, 
this question was addressed at one-loop order \cite{Belitsky:2024rwv}, where the tree-level double splitting was sufficient to find the 
integrand employing the unitary-cut sewing technique \cite{Bern:1994zx,Bern:1994cg,Bern:2004cz}. At two loops, in addition to
iterated double two-particle cuts, one has to include a three-particle cut of the five-leg superamplitude and the triple-collinear 
superamplitude found in this work.

These questions are currently under study and their results will be announced elsewhere.

\begin{acknowledgments}
The work of A.B. was supported by the U.S.\ National Science Foundation under grant No.\ PHY-2207138. 
The work of V.S. was conducted under the state assignment of Lomonosov Moscow State University.
\end{acknowledgments}

\appendix

\section{Equivalence of five-leg superamplitudes}
\label{AppendixEquivalence}

In this Appendix, we provide a simple analytical demonstration of the equivalence between the compact representation of the five-leg 
reduced superamplitude \re{CompactA5} to the original lengthier form \cite{Dennen:2009vk}
\begin{align}
\label{5legTree6D}
\widehat{\mathcal{A}}_5^{(0)} 
&
= \frac{1}{\prod_{i = 1}^5 S_{i, i+1}}
\Big(
Q_1 \bar{P}_2 P_3 \bar{P}_4 P_5 \bar{Q}_1 + \mbox{cyclic}
\\
&
+
\ft12
\left(
Q_1 \bar{P}_2 [P_3,\bar{P}_4,P_5] \bar{Q}_2
+
Q_3 \bar{P}_4 [P_5,\bar{P}_1,P_2] \bar{Q}_4
+
(Q_3 + Q_4) \bar{P}_5 [P_1,\bar{P}_2,P_3] \bar{Q}_5 + \mbox{cc}
\right)
\Big)
\, , \nonumber
\end{align}
where $[P_i, \bar{P}_j,P_k] \equiv P_i \bar{P}_j P_k - P_k \bar{P}_j P_i$. The $\Omega$ covariants in Eq.\ \re{CompactA5} are cyclically 
symmetric on the support of the super-momentum conserving delta functions,
\begin{align}
\label{CyclicOmega5}
\Omega
\equiv
\Omega_{12345} = \Omega_{23451} = \Omega_{34512} = \Omega_{45123} = \Omega_{51234}
\, .
\end{align}
We will focus on the $\Omega_{51234}$ ordering since it will be particularly useful for our follow-up study of the $4||5$ collinear 
asymptotics adopted in Sect.\ \ref{DoubleCollinearSection} of the body of the paper. Making use of the relations between super-region 
momenta and super-Poincar\'e ones \re{superPoincare2Dual}, we can recast the bras and kets \re{BconfCovs} entering 
$\Omega_{51234}$ as
\begin{align}
\bra{B_{5,13}} = - S_{34} Q_5 + (Q_3 + Q_5) (\bar{P}_3 + \bar{P}_4) P_5
\, , \quad
|\bar{B}_{5,24}]
=
- S_{51} \bar{Q}_4 + \bar{P}_4 (P_1 + P_5) (\bar{Q}_5 + \bar{Q}_1)
\, .
\end{align}
The inner products $\bra{B}\bar{B}]$ can be further simplified, to give
\begin{align}
\label{SimplifiedOmega}
- 2 \Omega_{51234}
&
=
S_{45}
\left[
S_{34} Q_5 \bar{Q}_1 - Q_4 \bar{P}_3 P_5 \bar{Q}_1 + S_{51} Q_3 \bar{Q}_4 - Q_3 \bar{P}_4 P_1 \bar{Q}_5
-
Q_3 \bar{P}_4 P_5 \bar{Q}_1
\right]
\\
&
+ S_{51} Q_4 \bar{P}_3 P_5 \bar{Q}_4 + S_{34} Q_5 \bar{P}_4 P_1 \bar{Q}_5
+
Q_4 [S_{34} S_{51} - \bar{P}_3 P_5 \bar{P}_4 P_1] \bar{Q}_5
+
{\rm cc}
\, . \nonumber
\end{align}

Now, to show its equivalence to \re{5legTree6D}, it is necessary to use super-momentum conservation conditions in 
Eq.\ \re{5legTree6D}, such that only the structures emerging in the dual conformal representation show up. It is not an easy 
fit, as a priori one could end up with multiple extra contributions that will vanish only upon the use of cyclicity and momentum 
conservation. One would want to avoid these unnecessary complications. The form of the invariant $\Omega_{5123}$ gives 
us a hint on how to proceed. Namely, the first term in it depends on the supercharges $(Q_3, Q_4, Q_5) \times (\bar{Q}_1, 
\bar{Q}_4, \bar{Q}_5)$. Thus, we will eliminate the rest from the amplitude \re{5legTree6D}. This is not completely unambiguous 
since one can trade these supercharges between the direct and chiral conjugate contributions. Therefore, it necessitates a 
trial-and-error procedure. We can demonstrate the sought-after equivalence with the following steps. First, we use the relation 
by applying (anti)chiral charge conservation in the first term of the second line of Eq.\ \re{5legTree6D}
\begin{align}
Q_1 \bar{P}_2 [P_3, \bar{P}_4, P_5] \bar{Q}_2
=
- Q_1 \bar{P}_2 P_3 \bar{P}_4 P_5 \bar{Q}_1 
&
- (Q_3 + Q_4 + Q_5) \bar{P}_2 P_3 \bar{P}_4 P_5 \bar{Q}_1
\\
&
- Q_1 \bar{P}_2 [P_3, \bar{P}_4, P_5] (\bar{Q}_3 + \bar{Q}_4 + \bar{Q}_5)
\, . \nonumber
\end{align}
Together with its chiral conjugate, it eliminates the first term in the first line. The last term above, i.e., involving $Q_1$ is
moved to the cc contribution but its conjugate is extracted from it: this term matches the pattern we are looking
for. The second and third terms in the first line in \re{5legTree6D} can be brought to the form 
\begin{align}
Q_2 \bar{P}_3 P_4 \bar{P}_5 P_1 \bar{Q}_2
&
=
- \ft{1}{2} (Q_3 + Q_4 + Q_5) \bar{P}_1 P_5 \bar{P}_4 P_3 ( \bar{Q}_1 + \bar{Q}_4 + \bar{Q}_5)
+
{\rm cc}
\, , \\
Q_3 \bar{P}_4 P_5 \bar{P}_1 P_2 \bar{Q}_3
&
=
- \ft{1}{2} Q_3 \bar{P}_4 P_5 \bar{P}_1 P_2 ( \bar{Q}_1 + \bar{Q}_4 + \bar{Q}_5)
+
{\rm cc}
\, ,
\end{align}
again using the super-momentum conservation.
The rest of the contributions in Eq.\ \re{5legTree6D} already possess the required structure and so they are just left
intact, however, the remaining two self-conjugate terms, proportional to $Q_i \dots \bar{Q}_i$, with $i = 4,5$, are then
split evenly between the direct contribution and its cc image.  Summarising this short consideration, we find the numerator of the 
five-leg amplitude \re{5legTree6D} to be
\begin{align}
2 \mbox{Numerator}[\widehat{\mathcal{A}}_5^{(0)}]
=
&
- (Q_3 + Q_4 + Q_5) \bar{P}_1 P_5 \bar{P}_4 P_3 ( \bar{Q}_1 + \bar{Q}_4 + \bar{Q}_5)
\\
&
- 
Q_3 \bar{P}_4 P_5 \bar{P}_1 P_2 ( \bar{Q}_1 + \bar{Q}_4 + \bar{Q}_5)
- 
(Q_3 + Q_4 + Q_5) \bar{P}_2 P_5 \bar{P}_4 P_3 \bar{Q}_1
\nonumber\\
&
+
(Q_3 + Q_4 + Q_5) [\bar{P}_5, P_4, \bar{P}_3 ] P_2 \bar{Q}_1
+ 
Q_3  \bar{P}_4 [ P_5, \bar{P}_1 , P_2 ] \bar{Q}_4
\nonumber\\
&
+ 
( Q_3 + Q_4) \bar{P}_5 [ P_1, \bar{P}_2 , P_3 ] \bar{Q}_5
+
Q_4 \bar{P}_5 P_1 \bar{P}_2 P_3 \bar{Q}_4
+
Q_5 \bar{P}_1 P_2 \bar{P}_3 P_4 \bar{Q}_5
+
\mbox{cc}
\, . \nonumber
\end{align}
Although this representation looks lengthy, its equivalence to $-2 \Omega_{51234}$ is now easy to demonstrate. It boils down to
the use of momentum conservation and repeated application of Dirac algebra. Let us show it for just one term as an example. 
We find after a chain of transformations
\begin{align}
2 \mbox{Numerator}[\widehat{\mathcal{A}}_5^{(0)}]_{Q_4 \bar{Q}_4}
&
=
- Q_4 \bar{P}_1 P_5 \bar{P}_4 P_3 \bar{Q}_4 + Q_4 \bar{P}_5 P_1 \bar{P}_2 P_3 \bar{Q}_4 + {\rm cc}
\\
&
=
- S_{34} Q_4 \bar{P}_1 P_5 \bar{Q}_4 - Q_4 \bar{P}_5 P_1 ( \bar{P}_4 + \bar{P}_5 ) P_3 \bar{Q}_4 + {\rm cc}
\nonumber\\
&
=
- 2 S_{34} Q_4 [\bar{P}_1 P_5 + \bar{P}_5 P_1] \bar{Q}_4 - 2 S_{51} Q_4 \bar{P}_5 P_3 \bar{Q}_4
\end{align}
where in the second line we anticommuted the last two Dirac matrices in the first term, imposing the on-shell condition
$P_4 \bar{Q}_4 = 0$ along the way. While in the second term, we used the on-shellness of lines $P_i \bar{P}_i = 0$ and 
the momentum conservation. Next, in the second term of the second line, we anticommuted $\bar{P}_4$ to the right and
$\bar{P}_5$ to get the anticipated expression in Eq.\ \re{SimplifiedOmega} since the first term of the third lines vanishes
by means of the Dirac algebra and $Q_4 \bar{Q}_4 = 0$. The rest can be done accordingly term by term. We thus prove
the equality
\begin{align}
\mbox{Numerator}[\widehat{\mathcal{A}}_5^{(0)}] = - \Omega_{51234}
\, .
\end{align}

\section{Scalar projections of $\Omega$'s}
\label{AppendixScalarOmegas}

We list here the scalar projections of the dual covariants $\Omega_i$. Imposing conditions \re{Q345zero} and \re{Q12solQ6}, they are
\begin{align}
2 S_{12} \Omega_1^{\rm sc} 
&
= S_{34} Q_6 (\bar{P}_4 + \bar{P}_5) P_3 \bar{P}_2 P_1 \bar{Q}_6 + \mbox{cc}
\, , \\
\Omega_2^{\rm sc} 
&
=
\Omega_3^{\rm sc} 
=
0
\, , \\
2 S_{12} \Omega_4^{\rm sc} 
&
= - S_{45} Q_6 \bar{P}_5 (P_6 + P_1) P_3 \bar{P}_2 P_1 \bar{Q}_6 + \mbox{cc}
\, , \\
2 S_{12}
\Omega_5^{\rm sc} 
&
= S_{345} Q_6 \bar{P}_1 P_2 (\bar{P}_6 + \bar{P}_1) P_5 \bar{Q}_6 + \mbox{cc}
\, , \\
2 S_{12}
\Omega_6^{\rm sc} 
&
= S_{345} Q_6 (\bar{P}_4 + \bar{P}_5) P_3 \bar{P}_2 P_1 \bar{Q}_6 + \mbox{cc}
\, .
\end{align}

\section{Scalar component of triple splitting}
\label{AppendixCollinearCheck}

In this appendix, we offer a check on the correctness of Eqs.\ \re{Omega1triple} -- \re{Omega6triple}. We chose the scalar
component, which is found by setting $Q_4 = Q_5 = \bar{Q}_4 = \bar{Q}_5 = 0$ and integrating out $Q_6$ and $\bar{Q}_6$
as in Eq.\ \re{Q6integrationOut}. In this manner, we find
\begin{align}
\vev{\vev{\Omega_1^{|||} \Omega_4^{|||}}}_6
&
=
z_4 z_6 S_{12} S_{23}^2 S_{45} \big[ (\bar{z}_6 - z_4) S_{456} - \bar{z}_4 S_{45} + \bar{z}_6 S_{56} \big]
\, , \\
\vev{\vev{\Omega_1^{|||} \Omega_5^{|||}}}_6
&
=
z_4 z_6 \bar{z}_6 S_{12}^2 S_{23}^2 \big[ (\bar{z}_6 - z_4) S_{456} - \bar{z}_4 S_{45} + \bar{z}_6 S_{56} \big]
\, , \\
\vev{\vev{\Omega_1^{|||} \Omega_6^{|||}}}_6
&
= 2 z_4 z_6 \bar{z}_6 S_{12}^2 S_{23}^2 \big[ \bar{z}_6 S_{456} - S_{45} \big]
\, , \\
\vev{\vev{\Omega_4^{|||} \Omega_5^{|||}}}_6
&
= 2 z_6 \bar{z}_6 (\bar{z}_6 - z_4) S_{12} S_{23}^2 S_{45} S_{56}
\, , \\
\vev{\vev{\Omega_4^{|||} \Omega_6^{|||}}}_6
&
= z_6 \bar{z}_6  S_{12} S_{23}^2 S_{45} \big[ (\bar{z}_6 - z_4) S_{456}  - \bar{z}_4 S_{45} + \bar{z}_6 S_{56} \big]
\, , \\
\vev{\vev{\Omega_5^{|||} \Omega_6^{|||}}}_6
&
= z_6 \bar{z}^2_6  S_{12}^2 S_{23}^2 \big[ (\bar{z}_6 - z_4) S_{456}  - \bar{z}_4 S_{45} + \bar{z}_6 S_{56} \big]
\, ,
\end{align}
with all other projections vanishing. Substituting these expressions into \re{N6numeratorFinal} and using the triple-collinear
asymptotics of the Mandelstam invariants \re{TripleMandelstamInv}, we find
\begin{align}
\vev{\vev{\mathcal{N}_6}}_6
=
S_{12} \bigg( \frac{z_4 z_6}{\bar{z}_4} 
&
+ \frac{S_{45}}{S_{456}} \bigg)
\vev{\vev{\Omega_1^{|||} \Omega_4^{|||}}}_6
- 
2 S_{45}
\vev{\vev{\Omega_1^{|||} \Omega_5^{|||}}}_6
+
\frac{S_{45} S_{56}}{S_{456}}
\vev{\vev{\Omega_1^{|||} \Omega_6^{|||}}}_6
\\
&
+
\frac{z_4}{\bar{z}_4} S_{12}
\vev{\vev{\Omega_4^{|||} \Omega_5^{|||}}}_6
- 2 z_4 S_{12}
\vev{\vev{\Omega_4^{|||} \Omega_6^{|||}}}_6
+
\frac{z_4}{\bar{z}_6} S_{45}
\vev{\vev{\Omega_5^{|||} \Omega_6^{|||}}}_6
\, , \nonumber
\end{align}
which after a little algebra gives the factorized form of the splitting amplitude's numerator
\begin{align}
\vev{\vev{\mathcal{N}_6}}_6/(S_{12} S_{23})^2
&
=
[\bar{z}_4 S_{45} + \bar{z}_6 S_{56} + (z_4 z_6 - 3 \bar{z}_4 \bar{z}_6) S_{456}]
\frac{z_4 z_6 S_{45}}{\bar{z}_4 S_{456}}\big[(\bar{z}_6 - z_4) S_{456} - \bar{z}_4 S_{45} \big]
\, .
\end{align}
Here, the first factor cancels the spurious pole in Eq.\ \re{TripleColSplittingSuperAmp}. The rest agree with the
collinear limit of the six-dimensional uplift of the scalar amplitude \re{4DscalarComponent}.

\end{document}